%% file: pionct_lt_th_arxiv.tex
\documentclass[aps,twocolumn,prl]{revtex4}

\usepackage{amsmath}
\usepackage{bm}
\usepackage{dcolumn}
\usepackage{graphicx}

\begin{document}

\title{Scaling study of the pion electroproduction cross sections and the pion form factor}

\author{
T.~Horn,$^{1, 2}$ 
X.~Qian,$^{3}$
J.~Arrington,$^{4}$ 
R.~Asaturyan,$^{5}$
F.~Benmokthar,$^{2}$
W.~Boeglin,$^{6}$
P.~Bosted,$^{1}$
A.~Bruell,$^{1}$
M.E.~Christy,$^{7}$ 
E.~Chudakov,$^{1}$
B.~Clasie,$^{8}$
M.M.~Dalton,$^{9}$
A.~Daniel,$^{10}$
D.~Day,$^{11}$
D.~Dutta,$^{3, 12}$
L.~El Fassi,$^{4}$ 
R.~Ent,$^{1}$ 
H.C.~Fenker,$^{1}$
J.~Ferrer,$^{13}$
N.~Fomin,$^{11}$
H.~Gao,$^{3}$
K.~Garrow,$^{14}$
D.~Gaskell,$^{1}$ 
C.~Gray,$^{9}$
G.M.~Huber,$^{15}$ 
M.K.~Jones,$^{1}$ 
N.~Kalantarians,$^{10}$
C.E.~Keppel,$^{1, 7}$
K.~Kramer,$^{3}$
Y.~Li,$^{10}$
Y.~Liang,$^{16}$
A.F.~Lung,$^{1}$
S.~Malace,$^{7}$
P.~Markowitz,$^{6}$
A.~Matsumura,$^{17}$
D.G.~Meekins,$^{1}$
T.~Mertens,$^{18}$ 
T.~Miyoshi,$^{10}$ 
H.~Mkrtchyan,$^{5}$ 
R.~Monson,$^{19}$
T.~Navasardyan,$^{5}$
G.~Niculescu,$^{13}$
I.~Niculescu,$^{13}$ 
Y.~Okayasu,$^{17}$ 
A.K.~Opper,$^{20}$
C.~Perdrisat,$^{21}$ 
V.~Punjabi,$^{22}$
A.W.~Rauf,$^{23}$
V.~Rodriguez,$^{10}$
D.~Rohe,$^{18}$
J.~Seely,$^{8}$
E.~Segbefia,$^{7}$
G.R.~Smith,$^{1}$ 
M.~Sumihama,$^{17}$
V.~Tadevosyan,$^{5}$ 
L.~Tang,$^{1, 7}$ 
V.~Tvaskis,$^{1, 7}$
A.~Villano,$^{24}$
W.F.~Vulcan,$^{1}$
F.R.~Wesselmann,$^{22}$ 
S.A.~Wood,$^{1}$ 
L.~Yuan,$^{7}$
X.C.~Zheng$^{4}$\\
}
\affiliation{$^{1}$ Thomas Jefferson National Accelerator Facility, Newport News, Virginia, 23606}
\affiliation{$^{2}$ Department of Physics, University of Maryland, College Park, Maryland, 20742}
\affiliation{$^{3}$ Triangle Universities Nuclear Laboratory, Duke University, Durham, North Carolina, 27708}
\affiliation{$^{4}$ Argonne National Laboratory, Argonne, Illinois, 60439}
\affiliation{$^{5}$ Yerevan Physics Institute, Yerevan, Armenia}
\affiliation{$^{6}$ Florida International University, University Park, Florida, 33199}
\affiliation{$^{7}$ Hampton University, Hampton, Virginia, 23668}
\affiliation{$^{8}$ Massachussets Institute of Technology, Cambridge, Massachusetts, 02139}
\affiliation{$^{9}$ University of the Witwatersrand, Johannesburg, South Africa}
\affiliation{$^{10}$ University of Houston, Houston, Texas, 77204}
\affiliation{$^{11}$ University of Virginia, Charlottesville, Virginia, 22904}
\affiliation{$^{12}$ Mississippi State University, Mississippi State, Mississippi, 39762}
\affiliation{$^{13}$ James Madison University, Harrisonburg, Virginia, 22807}
\affiliation{$^{14}$ TRIUMF, 4004 Wesbrook Vancouver, British Columbia, Canada}
\affiliation{$^{15}$ University of Regina, Regina, Saskatchewan, S4S-0A2, Canada}
\affiliation{$^{16}$ The American University, Washington, District of Columbia, 20016}
\affiliation{$^{17}$ Tohoku University, Sendai, Japan}
\affiliation{$^{18}$ Basel University, Basel, Switzerland}
\affiliation{$^{19}$ Central Michigan University, Mount Pleasant, Michigan, 48859}
\affiliation{$^{20}$ Ohio University, Athens, Ohio, 45071}
\affiliation{$^{21}$ College of William and Mary, Williamsburg, Virginia, 23187}
\affiliation{$^{22}$ Norfolk State University, Norfolk, Virginia, 23504}
\affiliation{$^{23}$ University of Manitoba, Winnipeg, Manitoba, R3T-2N2, Canada}
\affiliation{$^{24}$ Rensselaer Polytechnic Institute, Troy, New York, 12180}
\date{\today}

\begin{abstract}
The $^{1}$H($e,e^\prime \pi^+$)n cross section was measured for a range of four-momentum transfer up to $Q^2$=3.91 GeV$^2$ at values of the invariant mass, $W$, above the resonance region. The $Q^2$-dependence of the longitudinal component is consistent with the $Q^2$-scaling prediction for hard exclusive processes. This suggests that perturbative QCD concepts are applicable at rather low values of $Q^2$. Pion form factor results, while consistent with the $Q^2$-scaling prediction, are inconsistent in magnitude with perturbative QCD calculations. The extraction of Generalized Parton Distributions from hard exclusive processes assumes the dominance of the longitudinal term. However, transverse contributions to the cross section are still significant at $Q^2$=3.91 GeV$^2$.
\end{abstract}

\maketitle
Conventional pictures of the hadron, in which partons play the dominant role, predict a factorization of short-distance and long-distance physics. The factorization scale is the four-momentum transfer squared ($Q^2$). Earlier measurements of inclusive processes such as deep inelastic scattering (DIS), confirm that in the limit of large $Q^2$, at fixed values of the Bjorken variable $x_B$ (the longitudinal momentum fraction of the hadron carried by the parton in the infinite momentum frame), such processes can be viewed as scattering from individual partons within the hadronic system. A similar factorization of scales may be expected to apply to hard exclusive scattering and allow for using perturbative QCD (pQCD) processes in the description of hadrons.

A factorization theorem has been proven for longitudinally polarized photons in meson electroproduction~\cite{Coll97}. It states that for sufficiently large values of $Q^2$, at fixed $x_B$, and fixed momentum transfer to the nucleon, $-t$, the amplitude for hard exclusive reactions can be expressed in terms of a hard process, a distribution amplitude describing the formation of the final state meson, and Generalized Parton Distributions (GPDs), which encode the non-perturbative physics inside the nucleon. This is known as the ``handbag'' mechanism. Note that a single hard gluon is exchanged in the hard subprocess, in which a virtual photon couples to a single quark inside the nucleon. Though it is expected that the factorization theorem is valid for $Q^2 >$ 10 GeV$^2$, to date it is unclear whether it may already be approximately valid at moderately high values of $Q^2$ under certain conditions~\cite{Fran99}.

One of the predictions of the factorization theorem is that the dominant virtual photon polarization is longitudinal in the limit of large $Q^2$. The corresponding electroproduction cross section scales to leading order like $\sigma_L \sim Q^{-6}$ at fixed $-t$ and $x_B$. The contribution of transversely polarized photons is suppressed by an additional power of $1/Q$ in the amplitude. In the $Q^2$-scaling limit, pQCD describes the short distance process and GPDs provide access to the non-perturbative physics. The dominance of the longitudinal contribution by a factor of $1/Q$ is important because it contains the GPDs one would like to extract. Recent studies of $\rho^0$ and $\omega$ cross sections at the Thomas Jefferson National Accelerator Facility (Jefferson Lab) show relatively good agreement with the $Q^2$-scaling prediction for values of $Q^2$ between 1.5 and 5.0 GeV$^2$. However, considerable transverse contributions complicate the isolation of contributions from the handbag mechanism from these data~\cite{Had05, Mor05}.

QCD factorization for hard exclusive processes can be related to the pion form factor, $F_{\pi}$. In fact, if one replaces the GPD in the handbag mechanism by the nucleon-pion vertex and the pion distribution, one obtains $F_{\pi}$ in the pQCD formalism. The modified mechanism includes the same single hard gluon exchange as the handbag mechanism~\cite{Ber01}. As a result, pQCD calculations of $F_{\pi}$ should be consistent with the experimental data where factorization applies.  

Longitudinal/transverse (L-T) separated pion production cross sections are one way to test the predictions of the factorization theorem for hard exclusive processes. The measurement of $\sigma_L$ allows for tests of the $Q^2$-scaling prediction, and, at the same time, of the consistency of $F_{\pi}$ with the pQCD prediction. The size of transverse contributions can be quantified via $\sigma_T$. The interference terms, $\sigma_{LT}$ and $\sigma_{TT}$, may provide additional information on the applicability of the QCD factorization prediction. Based on the $1/Q$ suppression of transverse contributions in the amplitude, the corresponding power laws are 1/$Q$ and 1/$Q^2$ for $\sigma_{LT}$/$\sigma_L$ and $\sigma_{TT}$/$\sigma_L$, respectively. 

In this article we study QCD scaling using previously published $^{1}$H($e,e^\prime \pi^+$)n cross sections as well as newly available results from the pion transparency experiment (E01-107), which was carried out in Hall C at Jefferson Lab. Data were acquired at $Q^2$=2.15 and 3.91 GeV$^2$ for values of $-t$ ranging between 0.145 and 0.450 GeV$^2$, at a center of mass energy of $W$=2.2 GeV. Cross sections were obtained from a 4-cm hydrogen target at two different beam energies for each of the two $Q^2$ points. The electroproduced charged pions were detected in the High Momentum Spectrometer (HMS), while the scattered electrons were detected in the Short Orbit Spectrometer (SOS). 

Scattered electrons in the SOS were selected using a gas \v{C}erenkov detector. Positively charged pions were identified in coincidence in the HMS using an aerogel \v{C}erenkov detector with refractive index of 1.015 \cite{Asa05} in combination with a gas \v{C}erenkov detector filled with perfluorobutane at 0.956 atm. Any remaining contamination from real electron-proton coincidences was eliminated with a coincidence time cut of $\pm$1 ns. Background from the aluminum target cell walls, always less than a few percent, and random coincidences ($\sim$1\%) were subtracted from the charge normalized yields. The kinematic variables $Q^2, W$, and $t$ were reconstructed from the measured spectrometer quantities. A more detailed description of the experiment can be found in reference~\cite{Clasth06}.

The unpolarized pion electroproduction cross section can be written in terms of the electron lab frame solid angle, $d\Omega_e$, the scattered electron energy, $E^\prime$, and the pion solid angle, $d\Omega_{\pi}$, as 
\begin{equation}
   \frac{d^5 \sigma}{d \Omega_e  dE_e^\prime  d \Omega_{\pi}} = J\left(t,\phi \rightarrow \Omega_{\pi}\right) \Gamma_v \frac{d^2 \sigma}{dt d \phi},
\end{equation}
where $J\left(t,\phi \rightarrow \Omega_{\pi}\right)$ is the Jacobian of the transformation from $dtd\phi$ to $d\Omega_{\pi}$, $\phi$ is the azimuthal angle between the scattering and the reaction plane, and $\Gamma_v$=$\frac{\alpha}{2 \pi^2} \frac{E^\prime_e}{E_e} \frac{1}{Q^2} \frac{1}{1-\epsilon} \frac{W^2-M^2}{2 M}$ is the virtual photon flux. The virtual photon cross section can be decomposed into contributions from transversely and longitudinally polarized photons, and interference terms between the two polarization states,
\begin{eqnarray}
\label{eqn-unsep}
  2\pi \frac{d^2 \sigma}{dt d\phi} & = & \frac{d \sigma_T}{dt} + \epsilon  \frac{d \sigma_L}{dt} 
                                    +  \sqrt{2 \epsilon (1 + \epsilon)}  \frac{d \sigma_{LT}}{dt} cos \phi \\ \nonumber
                                   & + & \epsilon  \frac{d \sigma_{TT}}{dt} cos 2 \phi.
\end{eqnarray}
Here, $\epsilon=\left(1+2\frac{|{\bf q^2}|}{Q^2}\tan^2\frac{\theta}{2}\right)^{-1}$ is the virtual photon polarization, ${\bf q^2}$ is the square of the virtual photon three-momentum, and $\theta$ is the electron scattering angle. The individual components in equation~\ref{eqn-unsep} were determined from a simultaneous fit to the $\phi$-dependence of the measured cross sections, $\frac{d^2 \sigma}{dt d\phi}$, at two values of $\epsilon$.

The experimental cross sections are calculated by comparing the experimental yields to a Monte Carlo simulation of the experiment. To take into account variations of the cross section across the acceptance, the simulation uses a $^1$H($e,e^\prime \pi^+$)n model based on pion electroproduction data. The model was optimized in an iterative fitting procedure to match the $-t$- and $Q^2$-dependence of the data. The $W$-dependence of the data is well described by the empirical form $(W^2-M_p^2)^2$, where $M_p$ is the proton mass~\cite{Bra77}. The resulting model cross section is known to 10\%.

The uncorrelated systematic uncertainty, which is amplified by 1/$\Delta \epsilon$ in the L-T separation, is 0.6\% at $Q^2$=2.15 GeV$^2$ and 3.7\% at $Q^2$=3.91 GeV$^2$. The latter is dominated by the model dependence of the cross section.

At $Q^2$=2.15 GeV$^2$ the model dependence is $<$1\% as the data were analyzed at the same values of $W$ and $Q^2$ at high and low $\epsilon$. This was achieved by placing constraints to equalize the $W$/$Q^2$ phase space, and obtaining average central values from datasets over the overlapping region. At $Q^2$=3.91 GeV$^2$ the $W$/$Q^2$ kinematic coverage at high $\epsilon$ does not overlap with the one at low $\epsilon$, and does not allow for selecting the bin centers from a common set of values. An explicit correction (on the order of 4\%) was required to perform the L-T separation at common values of $W$ and $Q^2$. Other contributions to the systematic uncertainties are the correlated systematic uncertainty (3.7\%), which is dominated by radiative corrections and pion absorption and the $-t$-correlated systematic uncertainty (1.2\%). 
\begin{figure}
\begin{center}
\includegraphics[width=3.5in]{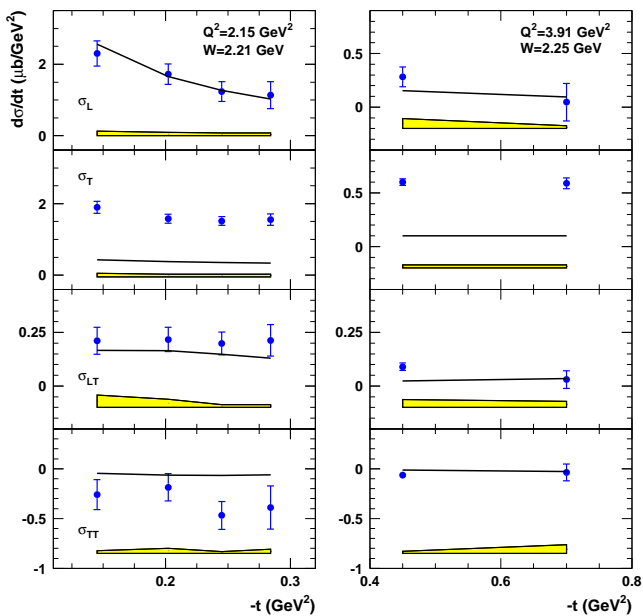}
\caption{\label{fig-sep-all} \it (Color online) The $-t$-dependence of the separated cross sections from E01-107 data. The error bars include the statistical and uncorrelated systematic uncertainty combined in quadrature and the error band includes the correlated and $-t$-correlated systematic uncertainties. The solid black lines denote the best fit of the VGL/Regge model~\cite{Van98} to the data. The fit parameter is the pion cutoff, which relates to the pion form factor in a monopole form. The fit value for the pion cutoff is $\Lambda_{\pi}^2$=0.518 at $Q^2$=2.15 GeV$^2$, and $\Lambda_{\pi}^2$=0.584 at $Q^2$=3.91 GeV$^2$. The value for the $\rho$ cutoff was taken to be $\Lambda_{\rho}^2$=1.7 at both values of $Q^2$.}
\end{center}
\end{figure}

Before examining the $Q^2$-dependence, the analysis was verified by comparison to the well known $-t$-dependence of the cross section. Our results are shown in Figure~\ref{fig-sep-all}. The separated cross sections were evaluated at fixed values of $W$, $Q^2$ and $-t$. At $Q^2$=2.15 GeV$^2$, $\sigma_L$ shows the characteristic exponential $-t$-dependence due to the pion pole, while $\sigma_T$, which is not dominated by the pion pole, is largely independent of $-t$. The interference terms are not large, but clearly contribute to the total cross section. Non-pole contributions become important as the kinematically allowed value of $t_{min}$ increases. At the highest $Q^2$, for which $-t_{min}$ is 23 times the pole value, $\sigma_L$ is independent of $-t$ within the uncertainty, while both $\sigma_{LT}$ and $\sigma_{TT}$ terms approach zero. 

An interesting observation is that the contribution of transversely polarized photons is still significant at $Q^2$=3.91 GeV$^2$. In fact, $\sigma_T$ is approximately a factor of two larger than $\sigma_L$. This trend may be expected since the pole dominated term, $\sigma_L$, decreases more rapidly with increasing distance from the pole ($-t$ $>$ 0.2 GeV$^2$), while $\sigma_T$ is largely independent of it. However, it should be noted that a significant transverse contribution to the cross section complicates the extraction of GPDs from the leading order term, which is assumed to be dominated by $\sigma_L$. 

The $Q^2$-dependence of $F_{\pi}$ extracted from the measured $-t$-dependence of $\sigma_L$ provides another way to verify the analysis. For the extraction a model developed by Vanderhaeghen, Guidal, and Laget (VGL) was used, which is based on the exchange of Regge trajectories. Most parameters are constrained by photoproduction data, leaving the squared mass scale of a pion monopole form factor at the electromagnetic vertices, $\Lambda_{\pi}^2$, as the only unconstrained parameter. By comparing the experimental value for $\sigma_L$ to the one from the Regge calculation, one obtains $F_{\pi}$ in a one parameter fit at each value of $Q^2$. This procedure is described in more detail in reference~\cite{Horn06}. 
\begin{figure}
\begin{center}
\includegraphics[width=3.5in]{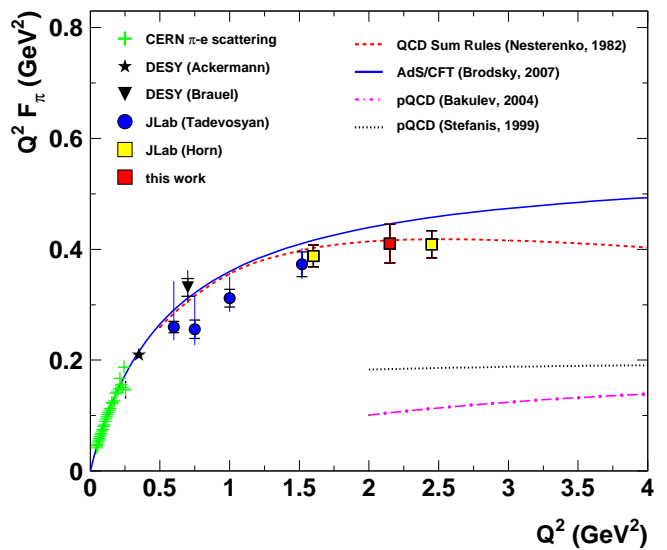}
\caption{\label{fig-fpi} \it (Color online) The pion form factor extracted at $Q^2$=2.15 GeV$^2$ shown together with previous data from CERN~\cite{Ame84}, DESY~\cite{Ack78,Bra77} and Hall C~\cite{Tad06, Horn06}. The extracted value for $Q^2 F_{\pi}$ is 0.410 $\pm$0.035, where the error bars do not include the model uncertainty.}
\end{center}
\end{figure}

The resulting value of $F_{\pi}$ extracted at $Q^2$=2.15 GeV$^2$ is shown in Figure~\ref{fig-fpi} along with data from CERN, DESY and Jefferson Lab~\cite{Ame84,Ack78,Bra77,Tad06, Horn06}. The new data point is in good agreement with the results from the latest Jefferson Lab experiment, which gives confidence in the reliability of the data analysis. The value of $-t_{min}$ at $Q^2$=2.15 GeV$^2$ is comparable to the $-t_{min}$ values used in the earlier Jefferson Lab measurements, while the $-t_{min}$ value at $Q^2$=3.91 GeV$^2$ is significantly larger ($t_{min} \sim$ 0.4 GeV$^2$). $F_{\pi}$ was not extracted at $Q^2$=3.91 GeV$^2$ due to insufficient knowledge of non-pole contributions in the extraction of $F_{\pi}$ at values of $-t_{min} >$0.2 GeV$^2$. 

For $Q^2 >$ 1 GeV$^2$, the $Q^2$ dependence of $F_{\pi}$ is consistent with the $Q^2$ power law behavior predicted by pQCD ($1/Q^2$)~\cite{Lep79}. On the other hand, the phenomenologically successful monopole fit (derived from vector meson dominace~\cite{Raj88}) would provide an equally good description of the data over the full kinematic range. The recently introduced anti-de Sitter space geometry/conformal field theory (AdS/CFT) correspondence is also consistent with the $Q^2$-scaling prediction, and also provides a reasonable description of the magnitude of $Q^2 F_{\pi}$ as shown in Figure~\ref{fig-fpi}. 

Perturbative QCD calculations, two of which are shown in Figure~\ref{fig-fpi}, give values of $Q^2 F_{\pi}$ between 0.10 and 0.20 GeV$^2$ in the region of our measurement~\cite{Bak04,Stef99}, significantly underpredicting the data. This may suggest that higher order contributions are still dominant at the available $Q^2$ scale~\cite{Mar98,Rad06,Nes82}. The broader shape of the AdS/CFT wavefunction, which contains the contribution from all scales up to the confinement scale, increases the magnitude of the leading twist pQCD prediction for the pion form factor by 16/9 compared to the prediction based on the asymptotic form~\cite{Choi07,Brod07}. However, it is still an open question how the AdS/CFT prediction of $Q^2 F_{\pi}$ asymptotically approaches the pQCD prediction~\cite{Choi07}.

\begin{figure}
\begin{center}
\includegraphics[width=3.5in]{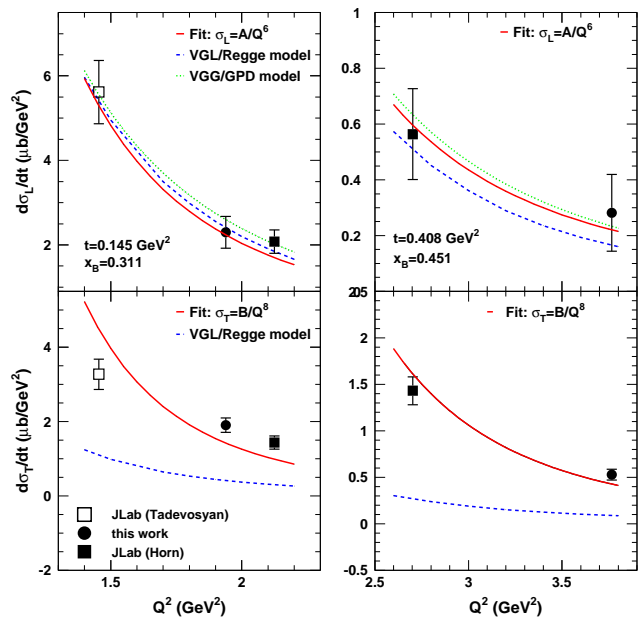}
\caption{\label{fig-fixed-xt} \it  (Color online) The Q$^2$-dependence of the separated cross sections at fixed values of $-t$ and $x_B$. The error bars denote the statistical and systematic uncertainties combined in quadrature. The red, solid curve shows a fit of the form $Q^{-6}$, for $\sigma_L$ and $Q^{-8}$ for $\sigma_T$.  The green dotted line is a GPD calculation from reference~\cite{Vdh98}. In this calculation power corrections to the leading order are included. The blue dashed line is a VGL/Regge~\cite{Van98} calculation using $\Lambda_{\pi}^2$ from a global fit to $F_{\pi}$.}
\end{center}
\end{figure}

The $Q^2$-dependences of the longitudinal and transverse cross sections, where results from this experiment have been combined with other recent results from Jefferson Lab~\cite{Tad06,Horn06}, are shown in Figure~\ref{fig-fixed-xt}. In order to compare the results at constant values of $-t$ and $x_B$, the results from the earlier Jefferson Lab experiments were scaled using the separated VGL/Regge cross section predictions. The additional uncertainty due to this kinematic scaling was determined from comparisons of the resulting cross section with one obtained using the GPD prediction (for $\sigma_L$ only) and a parameterization based on pion electroproduction data. The kinematic scaling uncertainty ranges between 7\% and 13\% in $\sigma_L$ for both $x_B$ kinematics, while the $\sigma_T$ uncertainty is larger by a factor of three. 

To examine one of the methods often used to test QCD factorization, we have compared our data to a GPD calculation by Vanderhaeghen, Guidal, and Guichon (VGG)~\cite{Vdh98}, which uses a factorized ansatz in terms of a form factor and a quark distribution function. The pion pole part of the amplitude is obtained using a parameterization of $F_{\pi}$ data including power corrections due to intrinsic transverse momenta and soft overlap contributions. The strong coupling constant between quarks and the exchanged gluon is deduced from the asymptotic freedom expression for $\alpha_s$ and by imposing $Q^2$ analyticity. This infrared (IR) finite form can be found in Refs.~\cite{Vdh98, Shir97}. It is interesting to note that both the Regge and the GPD models describe the $Q^2$-dependence of $\sigma_L$ quite well in this kinematic region. The agreement at low values of $-t$ may be expected as the Regge calculation, which is valid at low $-t$ and low $Q^2$, overlaps with the region of validity of the GPD calculation. However, this observation suggests that the agreement between data and GPD calculations is a necessary but not sufficient condition for the applicability of QCD factorization in this kinematic regime. 
 \begin{table}
 \begin{center}  % put inside center environment
  \begin{tabular}{|c|c|c|c|c|c|}
  \hline  % put a line under headers
  $x_B$  & $-t$  &  $\sigma_L \sim Q^{-n}$ & $\sigma_L \sim Q^{-6}$ & $\sigma_T \sim Q^{-m}$  & $\sigma_T \sim Q^{-8}$\\
         &       &  $n$                    & $\chi^2/\nu$ ($P$)     & $m$                     & $\chi^2/\nu$ ($P$)    \\
 \hline
 0.31 & 0.15   & 5.08 $\pm$ 0.95         &  0.45 (0.64)           &  4.20 $\pm$ 0.78        &  10.7 ($<$10$^{-3}$)      \\
 0.45 & 0.41   & 4.17 $\pm$ 2.95         &  0.24 (0.62)           &  6.01 $\pm$ 0.90        &  4.5  (0.034)      \\  
  \hline  % put a line under headers
  \end{tabular}
  \end{center}`
 \caption{\label{table_coeff} \it The central fit values for the $Q^2$ scaling laws $\sigma_L \sim Q^{-n}$ and $\sigma_T \sim Q^{-m}$ for both $x_B$ settings. Also shown are the $\chi^2$ values for fitting the data with the hard scattering predictions $\sigma_L \sim Q^{-6}$ and $\sigma_T \sim Q^{-8}$, and the corresponding probability, $P$, for finding this value of $\chi^2$ or larger in the sampling distribution assuming Poisson statistics.}
 \end{table}

A more stringent and model independent test of QCD factorization is the $Q^2$ power law scaling of the separated cross sections. The ``hard scattering'' predictions for $\sigma_L$ ($\sim Q^{-6}$) and $\sigma_T$ ($\sim Q^{-8}$) are indicated by the red lines in Figure~\ref{fig-fixed-xt}. To investigate the statistical impact of our data, we have fitted $\sigma_L$ and $\sigma_T$ at each value of $x_B$ to the forms $\sigma_L \sim Q^{n}$ and $\sigma_T \sim  Q^{m}$, where $n$ and $m$ are free parameters. The experimental fit values are listed in Table~\ref{table_coeff}. The best fit values for $\sigma_L$ are consistent with the hard scattering prediction within the uncertainty. In fact, fitting the $Q^2$-scaling prediction, $\sigma_L \sim Q^{-6}$, also results in a good description of the data. Note that the fit at $x_B$=0.451 is not well constrained due to the precision of the available data. While the scaling laws are reasonably consistent with the $Q^2$-dependence of the $\sigma_L$ data, they fail to describe the $Q^2$-dependence of the $\sigma_T$ data. The $Q^2$ dependence of $\sigma_T$ does, however, provide less conclusive evidence for having reached the hard scattering regime as the factorization theorem was proven rigorously only for longitudinal photons~\cite{Coll97}.

An interesting observation is that, at first glance, $\sigma_L$ appears to scale in a manner consistent with the $Q^{-6}$ hard scattering prediction, while the $F_{\pi}$ results in Figure~\ref{fig-fpi} obtained from these data are inconsistent with the magnitude of the two pQCD predictions shown. By definition of the hard scattering formalism, this would suggest that the QCD factorization regime has in fact not been reached. However, the discrepancy between $F_{\pi}$ data and pQCD predictions may arise from insufficient knowledge about how to extend the perturbative calculations to low values of $Q^2$. For example, recent calculations suggest that higher twist effects may still be significant up to values of $Q^2$ of 10-20 GeV$^2$~\cite{Rob98}. 

Higher order corrections play an important role at experimentally accessible energies~\cite{Goe01,Fran99}, and competing soft mechanisms may mimic the expected $Q^2$-scaling behavior characteristic for the hard pQCD term~\cite{Rad06}. Recall that the fitted scaling power for $\sigma_L$ could only be determined to $\pm$0.95, and the experimental uncertainties are rather large. In fact, deviations from the hard scattering prediction comparable to those shown in Table~\ref{table_coeff} have been observed in earlier measurements of the energy dependence of large angle Compton scattering performed at CEA~\cite{Shup79}. While the Compton data are in good agreement with the predicted energy dependence from pQCD, at least one theoretical model describes the data equally well, including the contribution of subleading logarithms at values of $Q^2<$ 10 GeV$^2$~\cite{Rad98}. Furthermore, recent data from Jefferson Lab~\cite{Dana07} covering the same kinematic range at higher precision show good agreement with the hard scattering prediction while the extracted scaling power at fixed center of mass angles differs considerably from the one predicted by pQCD.

It has been suggested that additional information about QCD factorization may be obtained through the interference terms~\cite{PR1206108}. However, the small size of these components may complicate the interpretation of the experimentally fitted scaling power. A fit to the interference terms from reference~\cite{Hornth06} suggests that the $Q^2$-dependence is reasonably well described by a functional form 1/$Q$ ($\chi^2$=0.94, probability=0.62) for $\sigma_{LT}$/$\sigma_L$, while a functional form of 1/$Q^2$ ($\chi^2$=1.34, probability=0.51) does a reasonable job describing the $Q^2$-dependence of $\sigma_{TT}/\sigma_L$ at $x_B$=0.311. However, the $Q^2$-dependence of the exponent can only be determined at the $\pm$17\% level for $\sigma_{LT}$.

In summary, separated pion electroproduction cross sections hold great promise in testing the factorization of long-distance and short-distance physics and the extraction of Generalized Parton Distributions from hard exclusive processes. The $Q^2$-dependence of separated $^{1}$H(e,e$^\prime \pi^+$)n longitudinal cross sections from Jefferson Lab are in relatively good agreement with the $Q^2$-scaling prediction. Recent results for $F_{\pi}$ including those presented here are also consistent with the $Q^2$-scaling prediction, but are not consistent in magnitude with pQCD calculations. The $Q^2$-scaling behavior is well reproduced by the AdS/CFT correspondence, which also provides a good description of the magnitude of $Q^2 F_{\pi}$. The asymptotic transition of the AdS/CFT to the pQCD prediction is not well understood yet. The extraction of Generalized Parton Distributions typically assumes the dominance of the longitudinal term. However, transverse contributions to the cross section are still significant at $Q^2$=3.91 GeV$^2$, and drop considerably more slowly than the $Q^{-8}$ scaling expectation. 

We thank S. Brodsky for helpful discussions. This work was supported in part by the U.S. Department of Energy. The Southeastern Universities Research Association (SURA) operates the Thomas Jefferson National Accelerator Facility for the United States Department of Energy under contract DE-AC05-84150. We acknowledge additional research grants from the U.S. National Science Foundation, and the Natural Sciences and Engineering Research Council of Canada.

\input{th-refs-pct}

\end{document}

%% file: th-refs-pct.tex
\hyphenation{Post-Script Sprin-ger}